\documentstyle[preprint,aps]{revtex}
\tightenlines
\begin{document}
\draft
\title{Final state interactions in the  $\eta d $ system}

\author{S. Wycech\thanks{e-mail "wycech@fuw.edu.pl"}}
\address{Soltan Institute for Nuclear Studies,Warsaw, Poland}
\author{A.M. Green\thanks{e-mail "anthony.green@helsinki.fi"}}
\address{Department of Physics and Helsinki Institute of Physics\\
P.O. Box 64, FIN--00014 University of Helsinki,Finland}
\date{\today}
\maketitle
\begin{abstract}
The $ \eta$-deuteron scattering amplitude is calculated with a recent K-matrix
model of $\eta N $ interactions. The existence of a  narrow
virtual state in the $\eta$-deuteron system is inferred. This state couples
strongly to the $ \eta d $ channel but its effect on final state interactions
in the $ pn \rightarrow \eta d$ reaction is shown to be rather weak.
Fairly large $\eta N $ scattering lengths are allowed by the experimental
data.
\end{abstract}

\newpage
\section{Introduction}
\label{intro}

Interactions of the $\eta$ meson with few-nucleon systems complement our
knowledge of the $\eta$-nucleon interaction.  An independent, although
related, interest in these systems stems from the conjectured existence of
$\eta$-nuclear quasi-bound states. Such states have
been predicted by Haider and Liu \cite{hai}, and Li et al. \cite{li},
when it was realised that the $\eta$-nucleon interaction is attractive.
So far there has been no direct experimental verification of this
hypothesis but, on the other hand, there is a mounting evidence that
such states exist in the lightest few-nucleon systems.

The earliest measurement was done in the four-body case. It was found at
SATURNE that the $ pd \rightarrow  \eta ^{3}$He production amplitude falls
rapidly just above the $\eta$ production threshold, \cite{gar},
\cite{mey96}. This led Wilkin to the suggestion that such a quasi-bound
state has been produced \cite{wil93}. A similar but weaker slope was found
in the $ dd \rightarrow \eta ^{4}$He amplitude  \cite{mey96},
and this may indicate that the quasi-bound states in five-body systems are
broader and located further away from the $\eta$ meson production
threshold \cite{hel}.

In more recent studies of three-body systems, very strong final state
interactions (FSI) were found in the  $ pp \rightarrow pp \eta$ cross
section in the threshold region \cite{cal},\cite{ber},\cite{chia}. Those led
to a speculation that this system may form a quasi-borromean state,
\cite{wycAPP}. As the final state interactions in the three-body continuum
are difficult to analyse, one finds perhaps  more convincing evidence
in the measurement performed at CELSIUS, \cite{pan97}. There,
the $\eta$-deuteron final states are observed in the scattering of a proton
on a deuteron target. This method allows the determination of the
$ pn \rightarrow \eta d$  amplitude, which turns out to be a smooth function
of energy but displays an enhancement close to the threshold.
Much  stronger enhancement is indicated by an  earlier SATURNE measurements
performed by  neutron scattering on a proton target \cite{plo}. These
data taken together indicate a
very sharp threshold effect in the $\eta$-deuteron  system. Earlier calculations
performed by Ueda confirm such an effect \cite{ued}, and indicate the
existence of quasi-bound   $\eta NN $ states.

The correlations seen in the three-body systems may be due to  quasi-bound
states or quasi-virtual states, i.e.three-body analogs of the deuteron or the
virtual spin singlet state in the $NN$ system.  However, there has been a serious
difficulty in such an interpretation. The strength of the $\eta N$
interaction and the magnitude of the scattering length $\ a_{\eta N}$ were not
large enough. The early values of Re $a_{\eta N}$
ranged between $0.25 fm $ and $0.5 fm $ ,\cite{hai},\cite{li},\cite{wil93},
\cite{kru},\cite{tia},\cite{ben},\cite{sau} although larger values have
also been suggested \cite{aba},\cite{ari}, \cite{bat},\cite{kai95},\cite{mos98}.
The lengths up to
$0.5 fm$ are not large enough to support long lived structures in
$\eta N N $ systems. Calculations have shown that the $\eta$-deuteron
virtual state is only likely to be formed provided  Re $a_{\eta N}$ exceeds a
critical value of about $0.7 fm$ ,\cite{gre96}. For larger values of about
$1.2 fm $, one finds the $\eta d$ system likely to be quasi-bound \cite{shev98}.
Such states would be reflected in a large value of the $\eta d$ scattering length
$ A_{\eta d}$. Unfortunately, the calculations of $ A_{\eta d}$
done in this critical region depend on details of the $\eta N$ interaction
model. The significance of the scattering length is one factor, but another
equally important one is the way the $\eta N$ scattering amplitude
extrapolates  to the region below the $\eta N$ threshold.

Recent phenomenological analyses of a four coupled-channel
($\eta N$, $\pi N$, $\gamma N$, $\pi \pi N)$ K-matrix model  have
helped to resolve these  difficulties.
In Ref. \cite{Kmatrix} a large scattering length $ a_{\eta N}=0.75(4)+i0.27(3)fm$
was found, and in addition the effective range expansion was shown to work well
in a broad region of both positive and negative energies. A very similar value of
$a_{\eta N}=0.72(3)+i0.26(2)fm$ was obtained in a dispersion theoretic
analysis of reference \cite{bat98}.
However, further studies \cite{Unc} of the K-matrix model
done in the spirit of Ref. \cite{Kmatrix} have
now revealed the existence of  another  solution, which  is characterised
by a small  scattering length of $ a_{\eta N} \approx  0.3 +i0.2 fm$ and
a very large effective range.
It should be added that a low value of $a$ is also produced by the GW21 solution
in the authors $N(1535)$  pole-position studies of Ref.\cite{Dick}.
\footnote{The authors wish to thank Bengt Karlsson for emphasising this point.}
In that work only
the channels $\pi N$, $\eta N$ and $\pi \pi N$ were treated explicitly. However,
when we use an extended version of the $K$-matrix model, in which both the
$\gamma \pi$ and $\gamma \eta $ channels are included\cite{Kmatrix1} then
this GW21 solution disappears leaving only GW11 with its large scattering length
Re $ a_{\eta N}=0.87fm$.
One hopes that the few-body data are going to be selective on these possibilities.
In addition, if $ a_{\eta N}$ is large there should arise a rich spectroscopy of
nuclear $\eta $ states. Forthcoming experiments may resolve this question,
\cite{hayano}, although difficulties in the interpretation have been
suggested, \cite{oset}.

It seems that now there exist enough information to study the $\eta N N $
phenomena in an almost quantitative way, and some related problems
are studied in this paper :

\noindent In section II, the $\eta$-deuteron scattering amplitude at low
energies is calculated. A method of partial summation of the multiple
scattering series used previously to calculate the eta-deuteron scattering
length,\cite{gre96},  is extended to the case of the continuum.
A strong cusp in the $\eta d $
scattering amplitude at the threshold is found and attributed to a
virtual state. The width of this state is very small, in the 1 MeV range.

\noindent In section III, the final state interactions in the $\eta d $ system
are discussed. We find the virtual state to couple rather weakly to this
system. It produces rather limited effects at energies below the deuteron
breakup. The results of Ueda,\cite{ued}, who has predicted a broad and
prominent peak above the $\eta d $ threshold, are not reproduced.

\noindent In section IV, we discuss the constraints induced  upon the
$\eta N $ scattering matrices by the $\eta NN $ data.
Also, the nature of  exotic $\eta NN $ states is discussed in this
section.

Some parts of this research could be,  and have been,  done
in an exact way by solving the three-body Faddeev  equations
\cite{shev98},\cite{del99},\cite{gar00}.
The solution for $\eta d $ scattering is the simplest one that
can be improved in this way.
However, even for this simple question the
general multiple channel analysis has never been fully implemented. It
 seems also to be premature, as many ingredients of the $\eta N $
interaction are not known. The advantage
of the method presented here  consists in the following :

(1) It is numerically simple and transparent in its physical interpretation.
Uncertainties in the two-body interactions and the meson formation process
may be separated and  discussed in simple terms.

(2) This method is applicable to more complicated few nucleon systems.

(3) It is very accurate numerically, at least in the calculation of the $\eta d$
scattering length.

\section{ The $\eta$ deuteron system at low energies}

We discuss the $\eta d$ scattering at very low energies.
This section contains three parts. First  a simple
method is introduced which allows us to find  the scattering matrix $ T_{\eta d}$.
At very low energies, below deuteron breakup, $ T_{\eta d}$ is dominated
by a narrow virtual three-body state. Next, we study the off-shell extension of
$ T_{\eta d}$ in order to construct the $\eta d $ wave function. The latter is
used to describe final state interactions in the $ \eta d $ formation processes.
In the third part the static nucleon approximation is discussed. Everywhere throughout this
paper the dominance of S waves is assumed.

\subsection{ A Formula for the $\eta$-deuteron T  matrix}

The purpose of this subsection is to derive a simple formula which relates
the low-energy meson-deuteron scattering amplitude to a few
multiple scattering integrals involving the meson-nucleon amplitudes.
The  formula  is  described in more detail in Ref.~\cite{gre96}, where
it was used to calculate the $ \eta d $ scattering length. Its numerical precision
was later  confirmed by a direct solution of this problem in terms of
the Faddeev equation \cite{del99}. In the limiting case of fixed nucleons
an agreement with other calculations has also been established \cite{rak}.
Here, this simple procedure is outlined  and its extension to the elastic or
inelastic continuum  and to the  off-shell region is discussed.

The starting point is a fixed nucleon
approximation. Next, upon the solution of this simple limit
a more realistic calculation is built up. It is achieved in a step-by-step
summation of the multiple scattering series until the necessary precision
is reached. The scattering length of a meson on a pair
of fixed nucleons is given by
\begin{equation}
\label{f1}
 A_{\eta d}=
\frac{2a_{\eta N} }{1-\frac{a_{\eta N}}{R_d}} ,
\end{equation}
where $a_{\eta N}$ is the meson-nucleon scattering length and $R_d$
is the inter-nucleon distance, \cite{bru}. Eq.(\ref{f1}) is obtained in a simple way
by setting boundary conditions for the meson wave function $\psi$ at each
scatterer as  $\psi '/\psi = 1/a_{\eta N}$. An alternative way to obtain
the same result  is to sum the multiple scattering series, which in this
case is a geometric series. With this expansion, the  $1/R_d$ term
stands for the meson propagator while $ a_{\eta N}\frac{1}{R_d} a_{\eta N} $
is a double scattering amplitude  etc.
The main advantage of Eq.(\ref{f1}) is that it offers the solution
even for a divergent multiple scattering series when
$ a_{\eta N}/R_d> 1$. The main weakness is that it is based on
distinguishable nucleons. Also, important kinematical and dynamical effects of
the three body propagation are missing and the actual value of $ R_d$ is left
undetermined.
The method to be described below retains the main advantage, but  removes
the weak points and offers a correct value of $ R_d$.

To improve Eq.(\ref{f1}) for the non-fixed nucleon situation  one needs to introduce:
the three-body propagator,
the  NN interactions and  an  off-shell $ \hat{t}_{\eta N} $ scattering matrix.
To achieve this, we study  the  multiple scattering expansion
for meson-deuteron scattering.
The basic $NN$ interactions are assumed to be known  and described by the deuteron
wave function  $\phi_d$ and a complete off-shell  $\hat{t}_{NN}$ scattering matrix.
With $\hat{t}_{NN}$ given one finds  a partial  three-body propagator

\begin{equation}
\label{f1a}
 G =   G_{0} +  G_{NN} ,
\end{equation}
where $G_0$ is  the free three-body propagator and
\begin{equation}
\label{f1b}
G_{NN} = G_0 \hat{t}_{NN} G_0
\end{equation}
constitutes that part of the three-body propagator which describes the $NN$ interactions.
Now, the series  appropriate for  the $\eta$-deuteron  scattering  matrix $\hat{T}_{\eta d}$
follows from the  Faddeev equations :
\begin{equation}
\label{f2}
\begin{array}{lll}
\hat{T}_{\eta d} & = &
\hat{t}_1+\hat{t}_2+\hat{t}_1G_0\hat{t}_2+\hat{t}_2G_0\hat{t}_1+\hat{t}_2G_0\hat{t}_1G_0\hat{t}_2+\hat{t}_1G_0\hat{t}_2G_0
\hat{t}_1  \\
& + & (\hat{t}_1+\hat{t}_2)G_{NN}(\hat{t}_1+\hat{t}_2)+(\hat{t}_1+\hat{t}_2)G_{NN}(\hat{t}_1+\hat{t}_2)G_{NN}
(\hat{t}_1+\hat{t}_2)+...,\\
\end{array}
\end{equation}
where the nucleons are labeled $1,2$ and the $\hat{t}_i$ are $\eta$-nucleon
scattering matrices.
This  expansion is performed in momentum space. All quantities involved in Eq.(\ref{f2})
are operators and this is denoted by the hat ($\hat{}$) symbols. Thus,  appropriate multifold
integrations over the intermediate Jacobi momenta ($\vec p ,\vec q$) must be performed.
In full generality the kernel of $\hat{T}_{\eta d}$ is a four point function
$T_{\eta d}(p,q,q',p';E)$  and to find the $\eta d$  scattering matrix
$T_{\eta d}$  a matrix element has to be calculated
\begin{equation}
\label{f3}
T_{\eta d}( p_i,E, p_f) =
<\phi_d \psi_{\eta}( p_i) \mid \hat{T}_{\eta d} \mid \phi_d \psi_{\eta}( p_f)>,
\end{equation}
where  the $\psi_\eta$ are  $S$-wave functions for the initial and
final  mesons of momenta $ p_i $, $p_f$.
In this section we are mainly concerned with  the  on-shell value of this
scattering matrix, which is denoted by $T_{\eta d}(E)$ with  $E$ being the
nonrelativistic   energy  of the $\eta d$ system.
The leading order (impulse  approximation) matrix  is
\begin{equation}
\label{f3a}
 T^0_{\eta d}  = <\hat{T}^0_{\eta d}> =  <\hat{t}_1 + \hat{t}_2 >,
\end{equation}
where $<>$ denotes the average over the deuteron and mesonic wave functions
as indicated in Eq.(\ref{f3}).
The  first  partial sum of the series  for $T_{\eta d}(E) $ is obtained with
the expression
\begin{equation}
\label{f5}
T^1_{\eta d}=   \frac{<\hat{T}^0_{\eta d}>}{1- \Sigma_1 - \Omega_1 }
\end{equation}
which contains the  double scattering terms
\begin{equation}
 \Sigma_1  =  \frac{<\hat{T}^0_{\eta d}G_{NN} \hat{T}^0_{\eta d}>}{<\hat{T}^0_{\eta d}>}
\label{f6}
\end{equation}
\begin{equation}
\label{f6a}
\Omega_1 = \frac{<\hat{t}_1 G_0 \hat{t}_2+ \hat{t}_2G_0 \hat{t}_1>}{<\hat{T}^0_{\eta d}>}.
\end{equation}
The first term  $\Sigma_1 $ describes the double scattering with an intermediate
NN interaction. It is  essentially an "optical model " double scattering term.
The second term $\Omega_1 $ includes  excitations of nucleons to  free continuum states.
The expression in Eq.(\ref{f5}) is  analogous to the one in Eq. (\ref{f1}),
since it contains terms linear in $<\hat{t}>$ in the denominator.
However, it also  now  contains
the prescription for
how to calculate the effective distance $R_d$  and additionally contains sizable effects
from the excitations.
Again, the expansion parameter is  $< \hat{t}>/R_d$ and it does not  need to be small
to guarantee the success of Eq.(\ref{f5}),  which is also applicable if the multiple
scattering series is divergent.
Higher orders of $<\hat{t}>/R_d$ in the denominator of Eq.(\ref{f5}) are obtained by comparing
higher orders in Eq.(\ref{f2}) with an expansion of Eq.(\ref{f5}) with respect
to $\Omega_1$ and $\Sigma_1$. In this way the next order  approximation is obtained as
\begin{equation}
\label{f7}
T^2_{\eta d} =\frac{<\hat{T}^0_{\eta d}>}{1- \Sigma_1 - \Omega_1 -S_2 }
\end{equation}
where
\begin{equation}
\label{f7a}
S_2 = [\Sigma_2-(\Sigma_1)^2]- [\Omega_2-(\Omega_1)^2]- [\Delta_2-\Omega_1\Sigma_1] \ \ ,\ \
 \Sigma_2=\frac{<\hat{T}^0_{\eta d}G_{NN}\hat{T}^0_{\eta d}G_{NN}\hat{T}^0_{\eta d}>}
{<\hat{T}^0_{\eta d}>},
\end{equation}

\begin{equation}
\label{f8}
\Omega_2=\frac{<\hat{t}_1G_0\hat{t}_2G_0 \hat{t}_1+\hat{t}_2G_0\hat{t}_1G_0\hat{t}_2 >}{<\hat{T}^0_{\eta d}>} \ \ , \ \
\Delta_2=2\frac{<\hat{T}^0_{\eta d}G_{NN}
(\hat{t}_1G_0\hat{t}_2+\hat{t}_2G_0\hat{t}_1)>}{<\hat{T}^0_{\eta d}>}.
\end{equation}

One may continue  this procedure to include higher
powers of $<\hat{t}>/R_d$ in the denominator, but for the present problem
the precision of the order of $1 \% $
has been already reached. This was shown in calculations of
the $\eta d $
scattering length,\cite{gre96},\cite{del99}, and optical model calculations for
$ \eta He $ systems, \cite{hel}. This fast convergence rate occurs as a
result of  an almost exact cancellation of the $\Sigma_2 - \Sigma_1^2$
and  higher order $\Sigma_n$ terms.
This  property holds also in the scattering region.

\subsection{Scattering amplitudes, off-shell extension, unitarity }

In this subsection some details of  the $\eta$ deuteron scattering
are discussed, the $\eta N$ interaction is introduced  and  explicit
expressions for the multiple scattering integrals are given.
Next, the question of unitarity is discussed for  energies
below the deuteron disintegration threshold.

The scattering matrices $T_{\eta d}(E),t_{\eta N}(E)$ used for the series summation
have the natural dimensions of length$^2$. To discuss the cross sections it is
more convenient to use the scattering amplitudes with the dimension of a length.
These are denoted  by $A_{\eta d}(E),a_{\eta N}(E)$, where  the convention
Im $A \geq 0$ is used
and the threshold values $ A_{\eta d}$, $a_{\eta N}$ are  the  scattering lengths.
The relations between scattering amplitudes and scattering  matrices  are
\begin{equation}
\label{f4}
A_{\eta d}(E) =- (2\pi)^2\mu_{\eta d}  T_{\eta d}(E),
\end{equation}

\begin{equation}
\label{f4a}
 a_{\eta N}(E) =- (2\pi)^2 \mu_{\eta N}  t_{\eta N}(E).
\end{equation}
For the multiple  scattering expansion, these relations introduce  factors
${\mu_{\eta d}}/{\mu_{\eta N}}$ which reflect  kinematical
differences in the propagation of these two systems.

Now, some definitions used in  this calculation are written down. The
integrations in Eqs.(\ref{f3})-(\ref{f8}) are performed in momentum space.
To describe wave functions the Jacobi coordinates $\vec q_{NN}$
(the relative $NN$ momentum)
and $\vec p_\eta$ ( the $\eta-NN $ relative momentum) are used.
However, to describe meson-nucleon interactions within the 3-body system
one uses other pairs  denoted by ($\vec q_{\eta N}, \vec p_N$).
A useful relation is $ \vec q_{NN} = \vec p_{N}- \vec p_\eta /2 $.
It allows one to express
the integration volumes  $d \vec p_{\eta} d \vec q_{NN}$ , $d \vec p_{N} d \vec q_{\eta N}$
also in terms of  $d\vec p_{N} d\vec p_{\eta}$.
The energy $E$ of the $\eta NN $ system is  related to the on-shell momentum $p_{\eta }$ by
$E= p_{\eta }^2 /2\mu_{\eta d} + E_d $ where $E_d $ is the deuteron
energy of $-2.2$MeV. For the free propagator one has now
\begin{equation}
\label{f4b}
G_0=[E_{NN}(q_{NN}) + E_{\eta}(p_{\eta}) -E]^{-1}
\delta (\vec p_{\eta}-\vec p'_{\eta}) \delta (\vec q_{NN}-\vec q'_{NN}).
\end{equation}

The meson-nucleon  scattering matrices are
$t_{\eta N} [q_{\eta N},q'_{\eta N},E-E(p_{N})] \delta
(\vec p_{N}-\vec p'_{N})$, where the delta-function is used to conserve
the spectator nucleon  momentum and $E(p_{N})$ accounts for the spectator nucleon
recoil energy. A typical on-shell $a_{\eta N}(E)$ is given in
Fig.1. The cusp in the real part at the $\eta N $ threshold
indicates an attraction that is due partly to the $N(1535)$ resonance
and partly to other attractive interactions.
It is this cusp  and the attraction related to it, that is responsible for
the attraction in the few-body-$\eta $  systems,
leading ultimately to the formation of  virtual or quasibound states.
The strength of the cusp i.e.  the value of scattering length
$a_{\eta N}$ is  controversial. The one shown in Fig.1 comes from
the phenomenological approach of Ref.\cite{Kmatrix},  but a number of
smaller as well as larger values have been predicted in the literature.
For this particular solution one finds the effective range expansion
\begin{equation}
\label{effr}
1/a_{\eta N}(E)=1/a_{\eta N} + \frac{1}{2} q_{\eta N}^2 r_{\eta N}
\end{equation}
with parameters  $ a_{\eta N}=0.75(4)+i0.27(3)fm $ and
$ r_{\eta N}=-1.50(13)-i 0.24(4)fm $
to cover the broad energy range in Fig.1.
Such a wide applicability of the effective range expansion
indicates  very short range $\eta N$ forces.
The attitude of this paper is to keep $a_{\eta N}$ as a semi-free parameter,
since it results from an interplay of the well established $N(1535)$  resonance
and unknown "background" interactions.
On the other hand,  the energy dependence of $a_{\eta N}(E)$ is determined
mostly by the $N(1535)$  which generates a negative effective range. Thus the value of
$ r_{\eta N}$ given above is believed to be weakly dependent on the
particular choice of $a_{\eta N}$.
Notice that $a_{\eta N}(E)$  is small at energies far below the threshold.
This allows a perturbative approach in this region.

Although $a_{\eta N}$ is "known" only on the energy-shell one also needs an
off-shell  extension. A separable one is used here:
$a_{\eta N} = v_\eta(q_{\eta N})a_{\eta N}(E)v_\eta(q'_{\eta N})$
with a Yamaguchi form factor $v_\eta =(1 + q_{\eta N}^2 / \kappa^2_\eta )^{-1}$.
The inverse range parameter $\kappa_{\eta}$ is apparently very large
for two reasons. Firstly, there is no long ranged meson exchange within
the ${\eta N}$ forces and, as said before, the effective range expansion converges rapidly.
Secondly, the form factors in any S-channel resonance must correspond
to the small sizes of these objects.
It has to be stressed
that in the $\eta d$ system at threshold one is close to a zero energy
binding situation. The results are sensitive to  every parameter involved
including the uncertain $\kappa_{\eta}$. A tentative value
$\kappa_{\eta}=3.3 fm^{-1},$
which corresponds to a natural radius of $N(1535),$  is used. The sensitivity
to $\kappa_{\eta}$, if it arises, will be indicated.
On the other hand, to simplify  notation all  formulas  are given in the
zero range $\eta N $  force limit. However, the actual  calculations are
performed without this restriction.

Within the three-body system, the $\eta N $ scattering matrix  is to be
averaged over some energy region, generated by the recoil of the spectator
nucleon.  So, the  $\eta d $
scattering length given by the impulse approximation
formula (\ref{f3a})  and normalisations  
in Eqs.~ (\ref{f4},\ref{f4a}) becomes
\begin{equation}
\label{f12}
A_{\eta d}^0(p_i,E, p_f)= 2 \frac{\mu_{\eta d}}{\mu_{\eta N} } < \int d\vec{p}
\phi_d(\vec{p}-\vec{p_i}/2) a_{\eta N}(E-\frac{p^2}{2\mu_{N,\eta N}})
\phi_d(\vec{p}-\vec{p_f}/2)>,
\end{equation}
where $ \vec{p_i} $ and $\vec{p_f}$ are the initial and final meson momenta.
Eq.(\ref{f12}) has been obtained with initial and final plane waves.
In order to extract the S wave contribution  an average over directions of
these momenta is to be performed and this is indicated by the brackets $<>$.
For on-shell scattering conditions one has $ p_i =p_f=p_{\eta }$.
This formula is more general, however, and
provides an off-shell extrapolation for $A_{\eta d}^0(p_i,E,p_f)$
to situations  where these quantities  are not related.
The integration in Eq.(\ref{f12}) averages the $ a_{\eta N}(E)$
amplitude over some region of recoil energies  $ p^2/2\mu_{N,\eta N}$.
The extent of this  region is determined by  momenta involved in the
deuteron wave function. At the threshold it starts with a negative $E_d$
and  extends towards negative energies down to about $-20$ MeV. For positive
energies, $E$   covers a range of about 20 MeV that includes the
$ \eta N$ threshold cusp.
For a limited region of $E$ it makes sense to introduce an
effective  amplitude  $\bar{a}_{\eta N}(E)$ that has been averaged
over these 20 MeV or so of the recoil
energies. We use Eq.(\ref{f12}) to define such an  average $\bar{a}_{\eta N}(E)$ and this
value will be used to calculate higher order scattering terms.
In this way the impulse approximation  amplitude becomes
\begin{equation}
\label{f12a}
A_{\eta d}^ 0(p_i,E,p_f)= 2 \frac{\mu_{\eta d}}{\mu_{\eta N} } \bar{a}_{\eta N}(E)
            < \int d\vec{p} \phi_d(\vec{p}-\vec{p_i}/2) \phi_d(\vec{p}-\vec{p_f}/2)>.
\end{equation}
It is expressed in terms of a  deuteron formfactor
\begin{equation}
\label{f12b}
F_d(p_i,p_f) =  < \int d\vec{p} \phi_d(\vec{p}-\vec{p_i}/2) \phi_d(\vec{p}-\vec{p_f}/2)> ,
\end{equation}
which   may be also presented in terms of space coordinates as
\begin{equation}
\label{f12d}
F_d(p_i,p_f) =
 < \int d\vec{r} \ exp[i\vec{r}(\vec{p_i}/2-\vec{p_f}/2)]  \phi^2_d(r)> =
 \int d\vec{r} \ j_0(rp_i/2)  \phi^2_d(r) j_0(rp_f/2).
\end{equation}
When the multiple scattering series is summed in Eq.(\ref{f7})
one obtains  the result
\begin{equation}
\label{f12c}
A_{\eta d}(p_i,E,p_f)=  \frac{ A_{\eta d}^0(p_i,E,p_f)}{1 -\Sigma_1 -\Omega_1 - S_2}.
\end{equation}
The effect of free continuum  $ \eta N N $  states is, to the lowest order
in $\bar{a}_{\eta N}$, given by $\Omega_1$.
As the  integrals in Eq.(\ref{f6}) are lengthy we reproduce this formula
only in the limit of zero range $\eta N$ forces
\begin{equation}
\label{f16}
 \Omega_1(p_{i},E,p_{f})=  \frac{ \bar{a}_{\eta N}}{  F_d(p_{i},p_{f})  }
\int \frac{d\vec{q} d\vec{q}'} {(2\pi)^2\mu_{\eta N}}
\frac{\phi_d(\vec{q}-\vec{p_i}/2) \phi_d(\vec{q}'-\vec{p_f}/2)}
{ [ E_{NN}(\frac{\vec{q}- \vec{q}'}{2})+E_{\eta}( \vec{q}+ \vec{q}')-E ]}.
\end{equation}
Now, this equation is used to define an inverse radius of propagation between
two successive collisions
\begin{equation}
\label{f16a}
 \Omega_1(p_{i},E,p_{f}) \equiv
 \bar{a}_{\eta N} \frac{\mu_{\eta d}} {\mu_{\eta N}} < \frac{1}{r}>_{3B}.
\end{equation}
Inverse radii analogous to $< \frac{1}{r}>_{3B}$
appear in many meson-deuteron  models, in particular in  pion-deuteron calculations.
Later we use these to compare  approximations and fix potential parameters.

To proceed any further, the nucleon-nucleon scattering matrix $t_{NN} (q, q', E) $ is
needed. It may be expressed by the NN scattering amplitude
via  a relation analogous to
Eq.(\ref{f4a}). To generate this matrix  a separable Yamaguchi potential
$V_{N N} = v_{NN}(q)\lambda_{N N} v_{NN}(q')$ is used with form factors
$v_{N N} = \kappa_{NN}^2 /( q^2 + \kappa_{NN}^2 )$. The two free parameters
$\kappa_{NN}$ and  $\lambda_{NN} $  are fixed  to reproduce
the deuteron binding and  one of the following two parameters: the  NN scattering length
or the inverse deuteron radius $< \frac{1}{r}>$. The details and 
motivation for the latter choice are presented later.
The  $t_{NN}$ enters multiple scattering series of the $\Sigma_n $ and $\Delta_n $ type.
As the relevant
integrals are lengthy we reproduce only the leading term in the limit of
zero range $\eta N$ forces and  energy averaged  $\bar{a}_{\eta N}$.
With this simplification
\begin{equation}
\label{f13}
\Sigma_1(p_{i},E,p_{f}) =  \frac{2\bar{a}_{\eta N}} {F_d(p_{i},p_f)}
\int \frac{d\vec{p}}{(2\pi)^2\mu_{\eta N}}
t_{NN}(E-\frac{p^2}{2\mu_{N,\eta N}})< V(\vec{p_{i}},\vec{p})  V(\vec{p},\vec{p_{f}})> ,
\end{equation}
where
\begin{equation}
\label{f14}
V (\vec{p}, \vec{p_{k}})=\int d \vec{q}
\frac{\phi_d(\vec{q}- \frac{1}{2}\vec{p_{k}})
v_{NN}(\vec{q}-\frac{1}{2}\vec{p})}
{E_{N}(\vec{q}-\frac{1}{2}\vec{p})+E_{\eta}(p)-E}.
\end{equation}
The angular average over $\vec{p_{i}},\vec{p_{f}}$
must be done to extract the S-wave contribution.
Eq.(\ref{f13}) defines another  inverse radius of propagation.
\begin{equation}
\label{f13h}
\Sigma_1(p_{i},E,p_{f}) \equiv
 2 \bar{a}_{\eta N}
\frac{\mu_{\eta d}} {\mu_{\eta N}} \ll \frac{1}{r}\gg_{3B}.
\end{equation}
The inverse radius  $\ll \frac{1}{r}\gg$ differs from   $< \frac{1}{r}>$
defined in Eq.(\ref{f16a}) above. Approximately, the former is  quadratic and the latter
is linear in the deuteron density and this is indicated by the notation.

The unitarity condition for the on-shell scattering amplitude  follows from
Eq.(\ref{f12c}). Let us concentrate on the energy region between the threshold and
the deuteron breakup. At low energies  $t_{NN}(E)$ is dominated by the deuteron
pole and it may be presented as
\begin{equation}
\label{f13a}
t_{NN} [E-E_{\eta }(p)] \approx  \frac {N_d^2}{E_d + E_{\eta }(p)-E },
\end{equation}
where $N_d$ is the deuteron wave function normalisation and $E_d$ is the deuteron binding
energy. The absorptive part of $t_{NN}$ comes from this singularity. Below the breakup
it  guarantees the unitarity for the  $\eta d $
scattering amplitude. To see that, one extracts the absorptive part of  $t_{NN}(E)$
equal to $i\pi N_d^2 \delta[E-E_{\eta }(p)- E_d] $
and performs the integration in Eq.(\ref{f13})  to obtain
\begin{equation}
\label{f13b}
Im \frac{\Sigma_1}{\bar{a}_{\eta d}}= 2 \frac {\mu_{\eta d}} {\mu_{\eta N}}
F_d(p_{\eta},p_{\eta }) p_{\eta }.
\end{equation}
The same procedure shows exact cancellation of the absorptive deuteron pole  contributions to
the triple scattering term $S_2$ in  Eq.(\ref{f7}). In addition, an inspection of  Eqs.(\ref{f6a}),
(\ref{f8}) shows that there is no pole in the integrands of the $\Omega_n $ terms.
In particular the singularity that enters $\Omega_1$ is related to the cut in $G_0$.
This  generates a branch point at the deuteron breakup threshold and absorptive
contributions at higher energies. Altogether, one can present the meson-deuteron scattering
amplitude in the form
\begin{equation}
\label{f11}
A_{\eta d}(E) = [\frac{1}{K_{\eta d}(E)}-ip_{\eta}]^{-1}
\end{equation}
as required by  unitarity.
Below the deuteron breakup threshold and for  closed $\pi N $ channels the
$K_{\eta d}(E)$ would be a real analytic  function of energy.
Above this  threshold, the $\eta N N$ continuum
induces an imaginary part to ${K_{\eta d}(E)}$.
It comes mostly from $\Omega_1$, much less from $\Sigma_1$ and effects of
higher orders are very small.

To describe final states in the $\eta d $ system a
half-off-shell scattering amplitude is  needed.
It is obtained by relating ${p_{f}}$ and $E$,  in the single and multiple scattering terms
$ F_d(p_{i},p_{f}(E)), \Sigma_1(p_{i},E,p_{f}(E)),  \Omega_1(p_{i},E,p_{f}(E))$ etc. In the
region of our main interest:  $E \approx E_d $  and $p_{i} $ in the
$0$ up to $ 1 fm^{-1}$ range  one finds
a  moderate (up to $ 25 \% $) fall of $ F_d(p_{i},0) $. On the other
hand the other quantities $ \Sigma_1(p_{i},0,E_d),  \Omega_1(p_{i},0,E_d)$ are almost  constant.
This  occurs since   these functions are defined by the ratios of two functions that fall down
with increasing  ${p_{i}}$ in the same moderate way. The consequence  is that the
half-off shell amplitude may be presented as 
I square some round brackets in next eq.
\begin{equation}
\label{ao1}
A_{\eta d}[p_{i},E,p_{f}(E)]
=\frac{A^0_{\eta d}[p_{i},E,p_{f}(E)]}{1-\Sigma_1(E)-\Omega_1(E)-S_2(E)}.
\end{equation}
For further use in the description of the $\eta d $ system in the process of $\eta d $ formation,
 Eq.~ (\ref{ao1}) is extended into an operator form
\begin{equation}
\label{ao2}
\hat{A}_{\eta d} =\frac{\hat{A}^0_{\eta d}}{1-\Sigma_1(E)-\Omega_1(E)-S_2(E)}.
\end{equation}
The  process of  formation and final state interactions requires more elaborate
calculations.
Approximations are done, and to study the value of these we first  discuss some
standard approximations used in elastic meson-deuteron scattering.

\subsection{The static nucleon and deuteron dominance approximations}

In the literature these approximations appear in many forms. The common element is
that intermediate
states of the nucleon-nucleon pair are reduced to the state of the deuteron.
Another, related approximation  neglects the recoil energy of the spectator nucleon.
Although  these approximations are rather crude  this method
is useful to  discuss limiting situations.  Below, it is used to indicate a
simple interpretation of the terms in our multiple scattering expansion.

The first double scattering contribution is the $ \Omega_1 $ term of Eq.(\ref{f16}).
If the nucleon recoil  $ E_{NN} $ is neglected, one obtains  at  $ E= 0 $

\begin{equation}
\label{st1}
\Omega_1^{static} =  \bar{a}_{\eta N}  \frac {\mu_{\eta d}}{\mu_{\eta N}}
             < \frac {1} {r} >,
\end{equation}
where  the  relevant "inverse deuteron radius" is
\begin{equation}
\label{st1a}
<\frac {1} {r}> =  \int  d\vec{r}   \frac {\phi_d(r)^2 }{ r}.
\end{equation}
The $< \frac {1} {r} > $ is a standard parameter in  low energy pion deuteron
scattering \cite{BEN00}. It describes the meson propagator at zero energy, averaged over the
deuteron density. One finds a weak but significant dependence of this parameter
on the NN potential used to calculate it.  In particular
$< \frac {1} {r} > = 0.448 /fm $ for the Paris NN potential
\cite{PAR81},  \cite{BEN00} and  $0.463 /fm $ for the Bonn potential \cite{BEN00}, \cite{BON87}.
In this calculation the  separable Yamaguchi potential is used which generates
the 
Hulth\'{e}n deuteron wave function. Two sets of potential parameters are tried. The standard choice for the potential strength and range reproduces  the deutron
binding and  $ pn  $  scattering length (Yamaguchi I). These are obtained with
$\kappa_{NN}= 1.41 fm^{-1} $ and $1 / \lambda_{NN}= -0.52 fm $. However, this choice generates
$< \frac {1} {r} > = 0.559 fm^{-1} $, which is too large  when  compared to
the superior calculations from the Bonn and Paris potentials.  To improve the separable model its  parameters are fixed
to reproduce the deuteron binding and  the  average of $< \frac {1} {r} > $ obtained
with these  two local potentials.
This requires $\kappa_{NN}= 0.91 fm^{-1} $ and 
$1 / \lambda_{NN}= -0.289fm $ (Yamaguchi II).
The $ np $ scattering length becomes $6.02 fm $ instead of  $5.40 fm $ but, for this
calculation,  the deuteron radius is more
important than the $ np $ scattering length.
The results are given in Table 1, where the
comparison with the exact three-body calculation is made.
A factor of two  difference is found. It comes from the large nucleon recoil
entering the propagator in  Eq.(\ref{f16}) and the deuteron  binding.
The latter is introduced into the three body calculation by the
condition  $ E= E_{d} $  ( not $ E= 0 $) at the threshold.

Now, we proceed to study the other double scattering term  $ \Sigma_1$.
At very low energies the deuteron dominance is expressed via
\begin{equation}
\label{g6}
G_{NN}(q,E,q') \approx \frac {\phi_d(q) \phi_d(q')}{E- E_d - E_\eta(p)}.
\end{equation}
This formula greatly simplifies all the  multiple scattering integrals.
To present the argument, the threshold scattering ($ E = E_{d}$)
is considered.  The dominant term   $ \Sigma_1 $  becomes
\begin{equation}
\label{st2}
\Sigma_1^{static} =  2\bar{a}_{\eta N}  \frac {\mu_{\eta d}}{\mu_{\eta N}}
            \ll \frac {1}{r} \gg ,
\end{equation}
where  another inverse deuteron radius  is introduced
\begin{equation}
\label{st3}
\ll \frac {1}{r} \gg =  \int  d\vec{r} d\vec{r'} \  \frac {\phi_d(r)^2 \phi_d(r')^2  }
{ \mid \vec{r} /2 -  \vec{r'}/2 \mid},
\end{equation}
in close analogy to the more general three-body definition of Eq.(\ref{f13h}).
The factor of  2 in the denominator is essential, since  $\Sigma_1^{static}$
corresponds  to the double scattering amplitude generated by an optical potential
\begin{equation}
\label{st4}
V^{static}(r) = - \frac {2 \pi }{\mu_{\eta N}}[2\bar{a}_{\eta N}]
[2^{3} \phi^2_d(2r) ].
\end{equation}
The strength of this  potential $2\bar{a}_{\eta N}$ is normalized to two nucleons.
The profile $\phi^2_d(2r)$ is "compressed" as it refers to the
center of the deuteron,
hence the argument $2r$ and the normalization factor $2^{3}$. $ V^{static}$
gives  the correct expression for the single scattering amplitude
but misses badly the double scattering term. In the static approximation
one obtains  $\ll1/r\gg  = 0.83 fm^{-1}$  while the  correct  three-body  value
 is $\ll1/r\gg_{3B}  = 0.32  fm^{-1}$
(Hulth\'{e}n-Yamaguchi I wave function and  zero range $\eta N$ force, see Table 1).
This large difference is, again, related to the nucleon recoil in the three-body
propagators. Now the propagator enters Eq.(\ref{f13}) twice and this makes the static
approximation even worse than in the $<1/r>$ case.

For scattering from a system of $A$ particles, a "remedy" for this
inconsistency has been found long ago by substituting
the $AA$ factor by $A(A-1)$.  For the deuteron this leads to  a
reduction of the  double scattering term by  one half.
One can  see in Table 1
that such a procedure brings the static optical potential
model  into reasonable consistency with the sum of $ \Sigma_1 $ and $ \Omega_1 $
terms.

\subsection{The pionic channels}

So far the effect of pionic channels has been hidden in the absorptive 
part of
the elastic $ \eta N$ scattering amplitudes. These describe processes related to
a single nucleon. In addition there exist processes  when an intermediate
$\pi$ meson is exchanged between two nucleons. The effect of such intermediate states is
expected to be small for two reasons. First,  the transition from
the $\eta$ to the $\pi$ meson involves a 300 MeV energy release. This energy is shared between
the three participants $N,N$ and $\pi$. The bulk of the   $NN$  interactions occur
at fairly high energies  where the $NN$ scattering amplitude is  small.
Second, at these energies the intermediate state propagator oscillates in space
at distances much shorter than the deuteron radius. This suppresses
the multiple scattering integrals.  However, since the $NN \eta$ system is close to
binding even small effects may matter. On  general grounds one could expect that
the  $NN \pi$ channel which has  a  lower threshold would contribute some repulsion into
the  $NN \eta$ channel which has  a  higher threshold. That could reduce the chances for binding in
the $NN \eta$ system.

With the  explicit pions  the series for the  $\eta$-deuteron  scattering  matrix $\hat{T}_{\eta d}$
is supplemented by terms
\begin{equation}
\label{f2p}
\Delta \hat{T}_{\eta d}  =
\hat{h}_1G_0\hat{h'}_2+\hat{h}_2G_0\hat{h'}_1 +
 (\hat{h}_1+\hat{h}_2)G_{NN}(\hat{h'}_1+\hat{h'}_2)+... \ ,
\end{equation}
where the scattering matrices $\hat{h}$ describe the  $\eta N \rightarrow \pi N$ transitions
and  the  $\hat{h'}$ describe  $\pi N \rightarrow \eta N$ transitions. These are related to the
scattering amplitudes via Eq.(\ref{f4a}). However, the kinematic factor is now
$ \sqrt{ \mu_{\eta N} \mu_{\pi N}}$. The small pion mass reduces the effect of  these
factors as  the relevant ratio $ \mu_{\pi N} /  \mu_{\pi d} $ is  close to unity.

The pion exchange scattering   is  described by
\begin{equation}
\label{f16p}
 \Omega_1^{\pi}(p_{i},E,p_{f}) =  \frac{\bar{a}_{\eta N,\pi N}^2 }{ \bar{a}_{\eta N} F_d(p_{i},p_{f})  }
I(m_{\pi}),
\end{equation}
where $I(m_{\pi}) $ has the form of the integral in Eq.(\ref{f16})
but with  the  mass of the $\eta$ meson
entering this integral being changed into the mass of  the  $\pi$ meson. The expression Eq.(\ref{f16p})
corresponds to the free propagator part of the series in Eq.(\ref{f2p}). The other terms are described
by a formula similar to Eq.(\ref{f13})
\begin{equation}
\label{f13api}
\Sigma_1^{\pi}(p_{i},E,p_{f}) =
\frac{2\bar{a}_{\eta N,\pi N}^2 }{ \bar{a}_{\eta N} F_d(p_{i},p_{f})  }
I(m_{\pi},m_{\pi}),
\end{equation}
where $I(m_{\pi},m_{\pi}) $ is given by the integral in
Eq.(\ref{f13}) but with the  mass of the  $\eta$ meson
being replaced by  the mass of  the $\pi$ meson.

A few  numbers calculated at the $\eta d$ threshold are now presented. These correspond to the
older version of the  $\eta N  $ model of Ref.\cite{Kmatrix}.
The scattering parameters  are :  $\bar{a}_{\eta N}=0.57+i0.14 fm  $ and
$\bar{a}_{\eta N,\pi N}=0.21+i0.15 fm  $,  while the double scattering integrals are
$I(m_{\pi})=(-3.5+i14.9)10^{-2} fm^{-1} $ and $I(m_{\pi},m_{\pi})=(6.89-i3.93)10^{-3} fm^{-1}$.
Thus the  $\Sigma_1^{\pi} $ is a negligible  1/pm correction to the multiple scattering. The
other term   $\Omega_1^{\pi}$ is significant.
Corresponding numbers for the elastic $\eta N $ channel may be found in Ref.\cite{gre96}
and in  Table I.

The calculation of $A_{\eta d}$ without the $\pi$ exchange effect
and the precision obtained with the multiple scattering theory  (MST)
can be estimated by comparing with the solution of  the  Faddeev equations (F)
obtained in Ref.\cite{del99} with the same input. For  small  $a_{\eta N} $
these methods give essentially the same answer.
At larger values, close to the binding situation  differences arise.
These are given in  Table II.

The MST  method slightly underestimates the real parts and overestimates imaginary parts.

The dominant
source of these differences is related to the use of energy averaged amplitudes
$\bar{a}_{\eta N} $ in the second and higher orders scattering terms. No attempt is
done here to correct for that.
Also there are higher order effects and  slight differences in the
effective range corrections involved in these two models.

Our best result for the $A_{\eta d}$  scattering length obtained with the
elastic $\eta N $ amplitude of Ref.\cite{Kmatrix} and the Yamaguchi II model for
the NN interactions is $A_{\eta d}= 2.31+i1.73 fm$.
This number should be compared with
$A_{\eta d}= 2.61+i1.72 fm $ obtained in Ref.\cite{gar00} from
the Faddeev equations, 
using the same $a_{\eta N} $ but a  somewhat superior separable $NN$ potential.
The results given above $do$ $ not$  include the pion exchange effects.
If the pion exchange is introduced we obtain
$A_{\eta d}= 2.27+i1.68 fm $. The change is rather small and
corresponds to a repulsive effect. Qualitatively, such a behaviour has been predicted in
Ref.\cite{gar00} but the size of the effect obtained there is
much larger and is not
reproduced here.
The difference may be related to the  small values 
of $\bar{a}_{\eta N,\pi N} $
inherent to the K-matrix  model of Ref.\cite{Kmatrix}.
 Eq.~ (\ref{f16p}) indicates that  $\Omega_1^{\pi}$ is composed of
three uncertain complex numbers and may be unstable with respect to
the input values of  $ \bar{a}_{\eta N,\pi N}$ 
and $ \bar{a}_{\eta N}$. 
Here, the averaged propagator
$I(m_{\pi})$ has been calculated with non-relativistic kinematics
as used in Ref.\cite{gar00}. A relativistic expression for the pion
energy would make the pion exchange effect even smaller.

Numerical results for the $\eta d$ scattering amplitude $A_{\eta d}(E)$
and the  $\eta d$ elastic cross section are given in Figs. 2a ,2b.
The impulse approximation is valid at energies $ E >1 5 MeV $
above the threshold. Multiple scattering effects enter at lower energies :
in the  intermediate region of $ E \approx 10 MeV $ it is
$\Omega_1$ that dominates while for lower energies  the main
contributor is $\Sigma_1$. The real three-body exotics is seen only very
close to the threshold for $ E <1 MeV + E_d $.
The effect is strong and very narrow. At present a direct way to see it
experimentally is not available, and one has to study inelastic processes
that involve the  $\eta d$ system in final states. One reaction of interest is
 $ pn \rightarrow d \eta$,  with a  related  one being
the $ pp \rightarrow pp \eta$.
In the next section  the deuteron reaction is discussed.

\section{ Final state interactions in the $\eta d $ system}

The amplitude squared for the $p n \rightarrow \eta d $  process extracted from
the CELSIUS experiment is given in  Fig.~\ref{fig4}. Three regions may be  specified :

(1) The "asymptotic" region of $E>20$MeV,  where the final state
interactions are small, can be described by the impulse approximation.

(2) The "interference" region of $E \approx 10$MeV,  where the experimental
amplitude indicates a  shallow minimum.

(3) The region of a "virtual three body state" located below the deuteron breakup
    threshold.

In this section we find  $ F_{\eta d}$,  the amplitude  for the $ pn \rightarrow d \eta $
reaction.  It consists of several terms that differ in their physical content.
In order to link this amplitude to the standard description of
 two-body  final state
interactions,  it is presented in the form
\begin{equation}
\label{fs1}
F_{\eta d}=  A^{c}[1+ \frac {A_{\eta d}(E)} {R_d(E)} ].
\end{equation}
This expression consists of the on-shell $\eta$-deuteron scattering amplitude
$A_{\eta d}(E)$ and a radius $R_{d}(E)$.
For simple models, like the static nucleon approximation,  a simple interpretation and
a closed formula for $ 1/ R_{d}(E)$ are found. The radius comes out as an
interplay of other radii: the $\eta $ meson source radius, the $\eta N$ force range
and  the deuteron radius. For scattering on static as well as interactive NN pairs
this radius  makes   sense only at very low energies.
At energies exceeding  about  10 MeV
it becomes strongly energy dependent. The presentation
of FSI in terms of  Eq.(\ref{fs1}) is still possible but the simple physical interpretation
is lost.

The three regions indicated in  Fig.~\ref{fig4} are specified for three different purposes.
The main emphasis is put  on the low energy enhancement and properties of the three-body  state.
There, it is the ratio $A_{\eta d}/ R_d  $ that represents the strength
of the enhancement.  In order to check on $A_{\eta d}$ one needs  full control
of  $ 1/ R_{d}$.
The interference region  indicates  an energy dependent
phase of $ 1/ R_{d}(E)$ that together with the phase of $ A_{\eta d}(E)$
produce a minimum. Finally the  cross section at higher energies is used
to fix   $A^c$ which is an  amplitude for coalescent formation of the deuteron.
To calculate it, one needs a specific model of the meson formation. This question
is discussed in the first subsection.

\subsection{ Meson formation amplitude }

The formation of an $\eta $  meson in nucleon-nucleon collisions involves
large momentum transfers  and the mechanism of
this process is uncertain. It is believed to be dominated by
an intermediate  $\rho$ meson exchange, although other mesons may also
contribute \cite{wil93},\cite{moalem},\cite{faldwil},\cite{batswe},
\cite{jul99}. In particular the recent calculation,\cite{RIS00}, stresses the
role of  $\eta$ meson exchange. All  models assume three stages for this
process, namely,  $NN \rightarrow NN(1535) \rightarrow NN\eta $, where the
intermediate
stage is dominated by the $N(1535)$ resonance. The first transition involves
large momentum transfers,  whereas    the second is a low energy process.

This paper aims at the final state effects and the formation
amplitude is described in a phenomenological way by a function
of initial and final momenta $A^{form}$.
What is essential for our study is to find a functional form  for 
this amplitude and pinpoint the effects (if any) that may generate
 a  rapid energy and  momentum dependence.
Such a dependence  would be equivalent to a long range space structure
of the $\eta $  meson source and this structure  is to  be viewed on  the
scale of the deuteron radius.

The $\eta$-meson production model is illustrated in 
Fig.~\ref{fig3}.  It consists
of a short ranged meson exchange part affected  by initial NN
interactions. As compared to the deuteron radius it is obviously a short range
process, since it requires large initial NN  c.m. momenta $Q_{NN} \approx 3.5 fm^{-1} $
and comparable momentum transfers. This part of the production process may
be described  by an amplitude $ ( q_t^{2} +  \mu ^{2} )^{-1}$,
where  $ q_t $ is the momentum transfer and $\mu$ is the mass of the exchanged meson.
Now, the momentum transfer  may be expressed in terms of the
initial and final momenta  $ \vec{ q_t}= -\vec{Q_{NN}}-\vec{p}_{1N}$,
where  the final nucleon  momentum is
$ \vec{p}_{1N}= \vec{q}_{NN}-\vec{p}_{\eta}/2$.
For the final states of interest, $ Q_{NN} \gg p_{1N} $ and
this amplitude is almost constant. Projected into  an
$S$ wave   it  generates a momentum  dependence
$ ( Q_{NN}^{2} +  \mu ^{2} +p_{1N}^{2} )^{-1} $.
As might have been guessed already from the uncertainty principle,
the range involved $ 1/ \sqrt{ Q_{NN}^{2} +  \mu ^{2} } $ is  less
than $0.35 fm $ even for   $ \pi$-meson exchange.

However, there is another part of the formation mechanism, related to the
propagation of the $N(1535)$  resonance relative to the spectator nucleon.
The propagation range is not negligible
on the scale of the deuteron radius. It may be described by a function
 \\ $G_{r}= [\Delta -i \Gamma(E)/2 +E_{rec}]^{-1} $ ,
where $\Delta = E_r -M_{N}-M_{\eta} - E$,
the nucleon recoil energy  $E_{rec}= p_{1N}^{2}/2 \mu_{\eta N,N}$,
and $E_r-i \Gamma(E)/2 $  are the resonance energy and width.
It is the recoil energy  that introduces a sizable momentum dependence
in this propagator and  generates a long range structure of the formation
process. Altogether, the formation amplitude is assumed to be of  the form

\begin{equation}
\label{sm}
A^{form}( p_{1N}) =
\frac{A^f 2 \pi^2}{\kappa_{s}^2 + ( p_{1N})^2},
\end{equation}
where $\kappa_{s}$ is the inverse radius of the $\eta$ formation source.
The Fourier transform of this amplitude with respect to the nucleon momentum
$ \vec{p}_{1N} $ becomes
\begin{equation}
\label{ss}
A^{form}(r) = \frac{A^f}{r } exp(- \kappa_{s}r).
\end{equation}
In general  $\kappa_{s}$, given by the relation
$ \kappa_{s}^{2}= 2 \mu_{\eta N,N}[\Delta -i \Gamma(E)/2 ]$,
is energy dependent but our main interest is the $ \eta N $ threshold.
The resonance position and half width of $1535-i148/2 $ MeV
are taken  from Ref.\cite{Kmatrix}. At the threshold, the decay of the
$N(1535)$ into  the $\eta N $  channel becomes negligible and
the actual width there is about $60$MeV.
These parameters yield a complex value of
$\kappa_{s}= 1.40 -i0.39 fm^{-1}$.
Strength of the meson formation amplitude $ A^{f}$ is a free
parameter. It is used to fix the calculated
$p n \rightarrow \eta d $  cross section to the experimental one
in the "asymptotic" region.

\subsection{ Final state wave function }

The full amplitude $ F_{\eta d} $ for reaction $ pn \rightarrow d \eta $
is given by an integral
\begin{equation}
\label{g2}
F_{\eta d} = \int  d\vec{q} d\vec{p} A^{form}(\vec{q}- \vec{p}/2)
\Phi^{-*}_{\eta d}(q,p),
\end{equation}
where the nucleon momentum  is  expressed in terms of Jacobi NN and $\eta $
momenta. In this equation $\Phi^{-}_{\eta d} $ is a wave function for the
final three-body
system that fulfills the ingoing wave boundary condition.
To find it, let us construct the multiple scattering series for the wave
function similar to the series for the scattering matrix. From the Faddeev equations one obtains
\begin{equation}
\label{g3}
\Phi_{\eta d} =
[1+ G_0 \hat{T}_{\eta d}+G_{NN} \hat{T}_{\eta d}] \phi_d \psi_{\eta} ,
\end{equation}
where $ \hat{T}_{\eta d}$ is  the scattering  matrix operator given by
the series in  Eq.~\ref{f2}. Now the reaction amplitude contains three distinctly
different terms
\begin{equation}
\label{g4}
F_{\eta d} = A^{c} +A^ {ex} + A ^{vs}
\end{equation}
that correspond to the three terms in the square bracket of Eq.(\ref{g3}).
The first one is a direct, coalescence transition from the source to the
deuteron :
\begin{equation}
\label{g5}
A^{c} = \int  d\vec{q} d\vec{p} A^{form}
(\vec{q}-\vec{p}/2)\phi_d(q) \psi_{\eta}(p).
\end{equation}
For an S-wave free meson final state of momentum $\vec{p}_{\eta}$ and the
specific choice of the formation amplitude in Eqs.(\ref{sm},\ref{ss}) the
coalescent amplitude is given by the integral
\begin{equation}
\label{g5a}
A^{c} = \int  d\vec{r} A^{form}(\vec{r})\phi_d(r) j_0(rp_{\eta}/2).
\end{equation}
The formation amplitude is short ranged in comparison to the deuteron radius.
Thus the coalescent formation amplitude involves $\phi_d(r \approx 0)$ and
in the region of interest it  depends weakly on  the meson momentum.


The second term in Eq.(\ref{g4}) describes a part of the $\eta$ 
meson final state interactions
\begin{equation}
\label{gex}
A^{ex} =
\int  d\vec{q} d\vec{p} A^{form}G_0(q,p) \hat{T}_{\eta d}\phi_d \psi_{\eta}.
\end{equation}
The superfix $ ex$ indicates that this term contains an  $\eta$ meson 
exchange diagram.  
Here the  $\eta NN$ system propagates freely from the formation source
and this process is described by $G_0$. Next, the scattering   series begins with the
meson-spectator-nucleon collision.
Thus the meson produced on one nucleon  is exchanged
and interacts with the other nucleon. After that the deuteron is built either in the
coalescent manner or in subsequent NN interactions.

The last term ($A ^{vs}$) in  Eq.(\ref{g4}) is  our main  interest. It describes the  final state
scattering process that begins with the  $NN$ interactions. These are the strongest
interactions in the  $\eta NN$ system. For low energies the  corresponding
propagator $G_{NN}$ is dominated by the state of the deuteron.
The related amplitude is
\begin{equation}
\label{g7}
A^{vs} = \int  d\vec{q} d\vec{p} A^{form}G_{NN}(q,p)
\hat{T}_{\eta d}\phi_d  \psi_{\eta}.
\end{equation}
On top of the deuteron described by $G_{NN}$ the system builts up the
$ \eta NN $ virtual  state ($vs$) and this process is described by $\hat{T}_{\eta d}$.
It is to be seen as a narrow enhancement on the background of the standard
coalescence term $A^{c}$.

Before discussing the $ \eta NN $ collective  state let us return to the meson
exchange process and the related amplitude $ A^{ex} $ of Eq.(\ref{gex}).
The final state scattering begins with the $\eta$-meson propagating from the source to the other
nucleon and in  leading order it is described by  $ G_{0}\hat{t}_{\eta N}$.
To this  order one can rearrange the scattering series in Eq.~(\ref{f2}) and the
final state wave  function to the form
\begin{equation}
\label{g3a}
\Phi_{\eta d} =
(1+ G_0 \hat{t}_{\eta n})[1+ G_{NN} \hat{T}_{\eta d}] \phi_d \psi_{\eta},
\end{equation}
which reduces the FSI to the coalescent  $ (A^{c}) $ and
 virtual state  $ (A^{vs}) $ amplitudes.
However, now the  the $ \eta $ formation
amplitude is redefined  to  $ A^{form}[1+ G_{0}\hat{t}_{\eta N}]$.
This  formation process is of an extended range in space
but is still much shorter than the deuteron radius.
The initial meson exchange process involves nucleons correlated
at short distances and thus  requires  large momenta and a 
large recoil. The relevant
energies in the $ \eta N $ system extend far below the
threshold.  There the $a_{\eta N}$ amplitude is not known. According to
the model shown in Fig.1 it is expected to be real and small.
Following this we terminate the exchange process at this level.
The result of this analysis is that the unknown strength of the
$\eta $  source is renormalized and its size may be somewhat larger
than that predicted by the $N(1535)$  resonance model.
In the numerical calculations a plausible value of  $\kappa_{s}= 1.5 fm^{-1}$
is used to describe the three factors that contribute to the size of the
$\eta $ meson formation source.

The task now is to find the coupling of the virtual state to the final
$\eta $ -  deuteron channel, and again the method of Sect.II is used 
for this purpose.
We begin with the simplest calculation.

\subsection{The deuteron dominance approximation}

This  approximation has been systematically used in all
calculations of the $\eta$  final state interactions on light nuclei.
In this section we discuss its applicability  in the deuteron
case. Ultimately it is shown that the use of  static nucleons is  a bad  approximation.
On the other hand,  this method is useful to discuss some approximations.
We take advantage of that.

At very low energies the deuteron dominance of $G_{NN}$ is expressed by
Eq.(\ref{g6}). This formula  inserted   into Eq.(\ref{g7}) gives  $A ^{vs}$
in terms the scattering matrix $ T_{\eta d}$.  One finds
\begin{equation}
\label{g8}
A^{vs} = \int d\vec{q} d\vec{p} \ A^{form}(\vec{q}- \vec{p}/2)\phi_d(q)
          \frac {T_{\eta d}(\vec{q},E,p_{\eta })}
         {E_d +E_\eta(p)-E}.
\end{equation}
The right hand side of this equation expresses $A^{vs} $ in terms of the
half-off-shell $ T_{\eta d}(\vec{q},E,p_{\eta })$ scattering matrix. The
problem that  arises at this stage is the off-shell continuation of
$ T_{\eta d}$. As discussed in the previous section the main dependence
of the  off-shell  extrapolation comes essentially from the numerator
of Eq.(\ref{f12a}) i.e. the deuteron formfactor of  Eq.(\ref{f12b}).
Within this approximation it is easy to extend the scattering to the
case of intermediate mesons which are described by plane waves.
The virtual-state amplitude in Eq.(~\ref{g8}) may be  presented in the coordinate
representation.
With the use of   the intermediate meson propagator
\begin{equation}
\label{g9}
\int d\vec{p} \frac {exp(i\vec{p}\vec{R})}
         {(2\pi)^2\mu_{\eta d} [E_d +E_\eta(p)-E]} = \frac{exp(i p_{\eta}R)}{R}
\end{equation}
and Eqs.(\ref{f12a},\ref{f12b}) one  obtains
\begin{equation}
\label{g8a}
A^{vs} =  \frac { A_{\eta d}(E) }{F_d(p_{\eta },p_{\eta }) }
          \int d\vec{r'} d\vec{r} \  A^{form}(r')\phi_d(r')
          \frac { exp[ i p_{\eta} \mid \vec{r}/2-\vec{r'}/2  \mid ] }{\mid \vec{r}/2-\vec{r'}/2  \mid  }
          \phi^2_d(r)  j_0(rp_{\eta}/2).
\end{equation}
The basic amplitudes $A^{c}$ and $A^{vs}$ are now easy to calculate in two
extreme cases : of a zero radius and a large radius meson  source.
For the zero radius source given by $A^{form}(r')= A^{f} \delta(\vec{r'})$ one obtains
the coalescence amplitude
\begin{equation}
\label{g9c}
A^{c} = A^{f}\phi_d(r=0),
\end{equation}
and the total amplitude  which  includes final state interactions
\begin{equation}
\label{g8b}
F_{\eta d}= A^{c}+ A^{vs} = A^{c}[1+ \frac {A_{\eta d}(E)} {R_d(E)} ],
\end{equation}
where the inverse interaction radius is defined as
\begin{equation}
\label{g10}
\frac {1} {R_d(E)} =
\frac{1}{F_d(p_{\eta},p_{\eta })}  \int d\vec{r} \phi^2_d(r) \frac{exp(ip_{\eta}r/2)}{r/2}   j_0(rp_{\eta}/2).
\end{equation}
With the zero range source and static nucleons  the  radius $R_{d}(E)$ is due
entirely to the deuteron size. For energies close to the threshold Eq.(\ref{g10})
relates it the previously defined static inverse radius
\begin{equation}
\label{g10a}
\frac {1} {R_d} = 2 < \frac {1}{r}> + i p_{\eta} + O(p_{\eta}^2).
\end{equation}
Numerically one obtains Re $ 1/ R_d  = 1.12 fm^{-1}$ for the Hulth\'{e}n I
deuteron wave function. The expansion  in Eq.(\ref{g10a}) adds a simple imaginary part
to  $ <\frac {1}{r}>$. For static nucleons  this is an exact expression as
may be  easily checked with Eq.(\ref{f12d}) for the deuteron formfactor.
For the FSI in  two-body systems  described by a potential, such an imaginary
term is due to the imaginary  part of  the outgoing wave $ \ exp(ipr)/r $.
It gives rise to an interference effect in the square-bracketed term of
Eq.(\ref{g8b}). By splitting the FSI amplitude into real and imaginary parts one
obtains
\begin{equation}
\label{g10b}
 \mid F_{\eta d} \mid^2  =  \mid  A^{c} \mid^2 \left[
[ 1+ Re A_{\eta d} <\frac {1}{r}> -p_{\eta} Im A_{\eta d} ]^2
 + [ Im A_{\eta d} <\frac {1}{r}> + p_{\eta} Re A_{\eta d} ]^2 \right].
\end{equation}
As the amplitudes in this equation are positive, one notices that,  with  increasing
momentum $ p_{\eta}$ and constant amplitudes,  the first term in Eq.(\ref{g10b}) decreases and the second
term increases. This leads to a minimum  in the amplitude. Such a minimum is indicated by the
experimental data in  Fig.~\ref{fig4}   in the region above the enhancement, where the
meson deuteron scattering amplitude becomes more stable.

The static nucleon approximation  is convenient   for the other simple limit of a large meson source.
If $A^{form}(r')= A^{f} \phi_d(r)$,  i.e. the source is as large as the
deuteron, then  the low energy expansion  is given by  formula (\ref{st3})
\begin{equation}
\label{g10c}
\frac{1}{R_d} =  \ll \frac{1}{r}\gg  + i p_{\eta}  + O(p_{\eta}^2).
\end{equation}
The radius $R_d$ given by this equation is sizably larger as one finds
$ \frac {1}{R_d} = 0.83 fm^{-1}$. The two examples given here indicate
the sensitivity of
the final state interactions to the source radius. For a more realistic source
determined  by the propagation of the $N^*$ resonance via Eq.(\ref{ss}) we find
$ Re \frac {1}{R_d} = 0.90 fm^{-1}$, at the threshold.

 However, static  nucleons are a rather poor approximation. The results are given in Table III, but
as may be guessed from Table I the
values of  $1 /R_d $ are likely to be overestimated by a factor of 2 to 3. To improve  this
situation it is necessary to analyze the dynamic deuteron formation
via the $NN $ interactions. This is studied in the next section.

\subsection{A dynamic formation of the deuteron}

The main approximation behind Eq.(\ref{g6}) is a reduction of three
body propagator to a quasi two body propagator. On the other hand, a correct expression
for the $NN$ propagator, that follows from the definition in Eq.(\ref{f1b}) is
\begin{equation}
\label{h1}
G_{NN}(q,q',E)= \frac{v_{NN}(q) t_{N N}[E-E_{\eta}(p)]v_{NN}(q')}
{[E_{NN}(q) + E_\eta(p) -E] [E_{NN}(q') + E_\eta(p) -E]}.
\end{equation}
For the threshold energy $E=E_d$  the denominators in the three body propagator
involved in this equation have no singularities. Together with the
form factors  $v_{NN}$ these generate the deuteron wave function
\begin{equation}
\label{h2}
\phi_d(q)= \frac {N_d  v_{NN}(q)} {[E_{NN}(q)- E_d]}
\end{equation}
provided the  energies $E_\eta(p)$ in the denominators  are  dropped.
The normalization factor
$N_d$ results from the residue of $t_{NN}$ at the deuteron pole.
Thus the static nucleon approximation of Eq.(\ref{h1})  consists in dropping
the $E_\eta(p)$ in the propagator  and reducing the $t_{NN}$ to a bare
singularity of Eq.(\ref{f13a}). These two approximations,  in part, tend
to compensate each other. 
Roughly,  the
two energies in the denominators $E_\eta(p)$ and $E_{NN}(q)$ are 
comparable 
two energies in the denominators $E_\eta(p)$ and $E_{NN}(q)$  are
comparable 
and the deuteron dominance  overestimates  $G_{0}^2 $ by a factor of  $2^2$.
This happens because the deuteron is a weakly bound object and the binding energies
in the denominators are small in comparison to the recoil energies.
On the other hand  by limiting  $t_{NN}$ to the pole term
one neglects a sizable attraction that exists in the $NN$
system at large negative energies. The pure pole term   of Eq.(\ref{f13a}) underestimates
$t_{NN}$ by a factor of  $1/2 $.  The net effect for $G_{NN}$ is  a reduction to about a half of
its static nucleon values. The details depend on the intermediate
momenta involved in this process.

In the dynamic approach the virtual state amplitude $A^{vs}$ is given by
an up-dated version of Eq.(\ref{g8}), where now the deuteron pole expression
for $G_{NN}$ given by Eq.(\ref{g6})
is substituted by the full expression of  Eq.(\ref{h1}). In addition, the half-off-shell
$\eta$ deuteron scattering amplitude given by Eq.(\ref{ao1})
is used. The full $A^{vs}$  is given by  an integral similar to
the integral in Eq.(\ref{f13})  for the $\Sigma_1$. However, in
Eqs.(\ref{f13}),(\ref{f14}) one of the deuteron wave
functions $\phi_d(\vec{q}- \frac{1}{2}\vec{p_{i}}) $ should be replaced by the source function
$A^{form}(\vec{q}- \frac{1}{2}\vec{p})$.
In the limit of the deutron domination this procedure returns back to the simple equation (\ref{g8}).
If one wishes to present $A^{vs}$ in the form of
Eq.(\ref{fs1}), then the inverse radius
$ \frac {1}{R_d}$ becomes a ratio of two integrals.
At the threshold, the effects of the three body propagator and 
source size reduce this quantity to
$ \frac {1}{R_d(0)}=0.46$.
This number is more than a factor of 2 smaller than that
given in the zero range source, deuteron dominance model. Some other values calculated for
two NN potentials and  two source ranges are given in Table III. It is also found
that the energy dependence of $R_d(E)$ is more moderate than  that given
by Eq.(\ref{g10}). This reflects the significance of the attraction in  $t_{NN}$
at  negative energies below the deuteron pole.

\section{RESULTS}

The  elastic $\eta d $ cross section is plotted in Fig. 2a for two
values of the inverse scattering parameters $ < \frac {1} {r} > $.
These parameters differ by $ 15\% $ only,  but the effect on the cross
section is significant. It reflects proximity of the  $\eta d $
quasibound state. A similar sensitivity  is displayed by the cross section  on
the value of the  $ \eta N $ scattering lengths. A few possibilities
are plotted in Fig. 2b.
These differ by the input scattering lengths $ a_{\eta N} $ and effective ranges.
Since the cross section is not measured,
it cannot be a test for   $ a_{\eta N} $. The point of interest is the relation of the peak
in the cross section to the position of the quasibound singularity in the complex
$ p_{\eta }$ plane.
This singularity is  found  as the position of zero in the denominator of Eq.(\ref{f12c}).
An approximate solution for $ a_{\eta N}=0.75+i0.27 fm $ case is
$ p_{vs}= -0.09-i0.09 fm^{-1}$ and it is located in the
third quadrant of the complex plane.
This corresponds to a virtual-state, that is  analogous 
 to the spin singlet proton-neutron state. Coupling to the pionic channels pushes
the singularity from the negative imaginary semi-axis into the third quadrant of the
$ p_{\eta }$ plane.  This singularity is far from the region where
quasi-bound  $\eta d$ states might exist.
Those would be  analogous to  the deuteron.
To enter the quasi-bound state region with the model discussed here, in particular with the
subthreshold amplitudes shown in Fig.1,  one would need 
$Re \  a_{\eta N} > 1.2 fm $.
The K-matrix model \cite{Kmatrix1} based on two body data 
allows scattering lengths larger than  $0.75 fm $, up to about 
$1.05 fm$. The corresponding  $\eta-d$ amplitudes are given in 
Fig.2b.    
However, as discussed below the large values close to $1 fm$ seem to be
excluded by the $ \eta d$ final state interactions.

In  Fig.~\ref{fig4} the CELSIUS data \cite{pan97} are plotted  
and compared with calculations.
The cross section around the 10 MeV excess energy region is used to fix 
our normalisation constant $A^f$.
The choice of fixing point is not particularly relevant, provided
it is done above the low few-MeV energy region. However, as is seen from
the figure,   our calculations are on the lower side of the experiment for
$E > 15 MeV $ and on the upper side for $ E < 3 MeV$. The 9 MeV point
chosen helps to make a good  overall  fit to the data.
At low energies the experimental energy resolution is vital.
The experimental points refer to the middle of the corresponding
energy bins. No attempt is made
to fold the calculated cross section  over the experimental
energy resolution.
 At low energies and in the enhancement region this cross section  is
a check on  the $ \eta N $ model.
The data and the results calculated with the  Yamaguchi II NN 
potential and
$ \eta N $ interactions of Ref.\cite{Kmatrix} are consistent. The same
data with the experimental errors enlarged by the beam energy uncertainty
(0.3 - 0.5 MeV)  could also accomodate  the results obtained
with the Yamaguchi I NN potential.  Large values of  $ a_{\eta N} $
would not be allowed,
at least with the subthreshold behavior generated by this K-matrix model.

The position of  the minimum in the scaled cross section comes out as an  interference effect.
It depends on  details of the energy dependent  $ \eta d$ scattering amplitude.
However, with the present precision of the data and uncertainties of the input,
it does not present a  test for the consistency of the data and
calculations.

{\bf Acknowledgements}
One of the authors (S.W) wishes to acknowledge the hospitality of the Helsinki University Department
of Physics and the Helsinki Institute of Physics, where part of this work was carried out.
This project is partially financed by the Academy of Finland under
contract 43982. We wish to thank Hans Cal\'{e}n, Andrzej Deloff and Joanna Stepaniak
for helpful discussions.

\begin{table}
\caption{Inverse deuteron radii ( propagator at zero energy averaged over deuteron wave functions)
calculated with  different deuteron wave functions
and/or different  propagators   for the $ \eta NN $ system. The $<>$ entries correspond
to free continuum  states: $ <> $  static nucleons, $ <>_{3B} $ 3-body propagator.
The $\ll\gg$  entries  correspond to intermediate NN interactions :
$\ll\gg$ projection  onto 
the deuteron state, $ \ll\gg_{3B} $ complete  NN interactions.
The three-body entries
{\em do not} include effects of a finite $ \eta N $ force range.
With  $ \kappa _{\eta N} = 3.3 fm^{-1}$
these should be reduced: $ <>_{3B}$ by  18 $\% $  and $\ll\gg_{3B}$
by  13$\% $. All entries are in $fm^{-1}$ }

\begin{tabular}{lcccl}
$                        $  & Paris       &  Bonn  &   Hulth\'{e}n II&  Hulth\'{e}n I  \\
$ < \frac {1} {r} >      $  & 0.449       & 0.463  &   0.456     &  0.559      \\
$< \frac {1} {r} >_{3B}  $  &             &        &   0.197     &  0.254      \\
$\ll \frac {1} {r} \gg     $  &             &        &   0.721     &  0.831      \\
$\ll\frac {1} {r}\gg_{3B}  $  &             &        &   0.280     &  0.324      \\
\end{tabular}
\label{table1}
\end{table}

\begin{table}
\caption{Comparison between the Multiple Scattering Theory (MST) used in
this paper and the Faddeev(F) results  for $A_{\eta d}$ -- see
Ref.~\protect\cite{del99}}
\begin{tabular}{lcl}
$a_{\eta N}$(fm)&$A_{\eta d}$(fm)(MST)&$A_{\eta d}$(fm)(F)\\
0.44+i0.30&1.01+i1.50&1.03+i1.49\\
0.62+i0.30&1.65+i2.41&1.71+i2.34\\
0.888+i0.274&2.37+i5.79&2.65+i5.48\\
\end{tabular}
\label{tableDeloff}
\end{table}

\begin{table}
\caption{Inverse radii $ < \frac {1} {R_d} >  $
used for the description of $ \eta d $ final state interactions at
zero energies.
All entries are in $fm^{-1}$}
\begin{tabular}{cccl}
Method & $\eta $ source  &  Hulth\'{e}n II &  Hulth\'{e}n I  \\
 Deuteron Dominance  & Zero range          &   0.91      &  1.12       \\
  "  & Resonance dominated        &   0.75      &  0.90       \\
 Deuteron Formation &  Zero range          &   0.504     &  0.555      \\
  "   &  Resonance dominated        &   0.465     &  0.528      \\
\end{tabular}
\label{table3}
\end{table}

\begin{figure}[ht]
\includegraphics{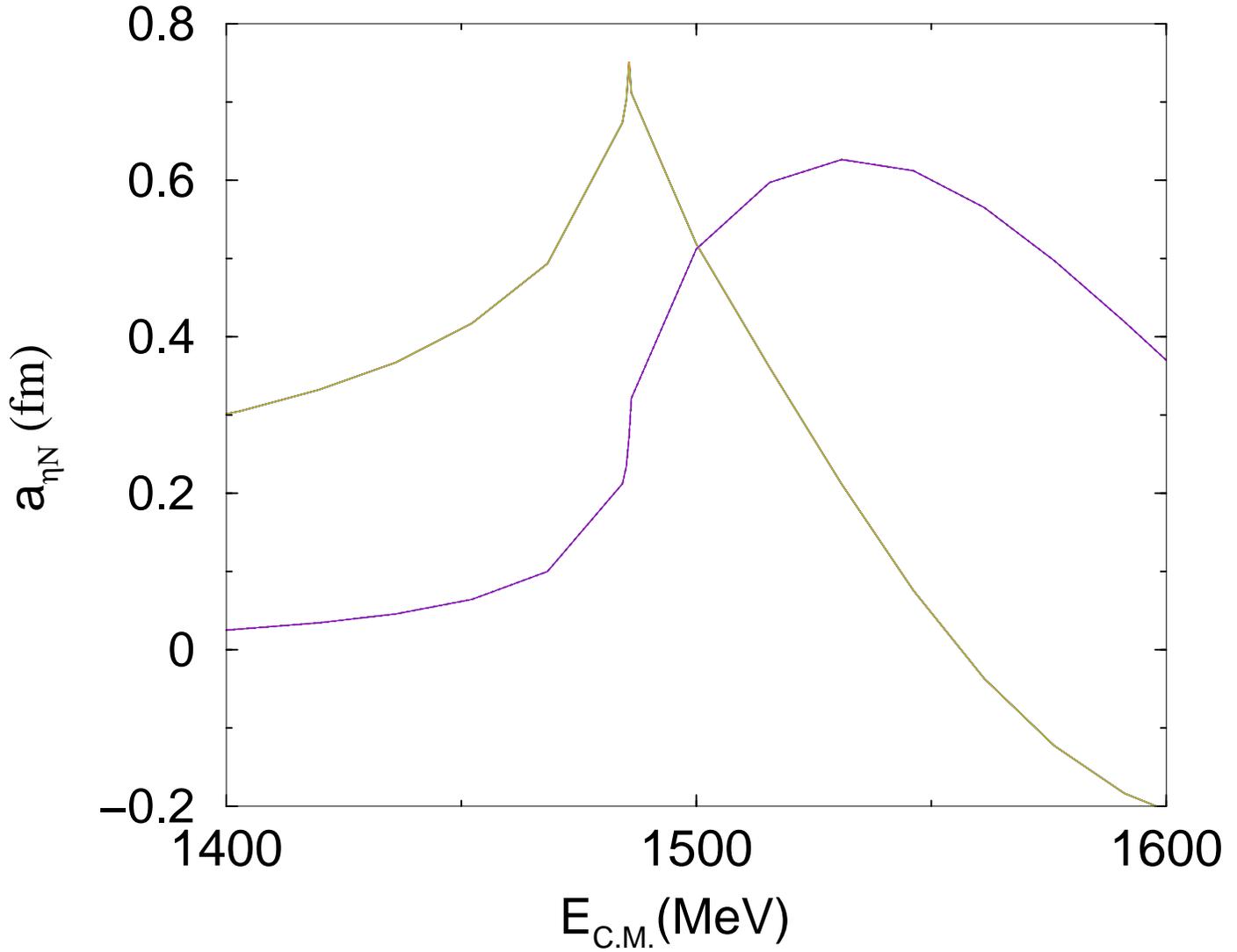}
\caption{The $\eta N\rightarrow \eta N$ amplitude  $a_{\eta N}$
 in fm as a function of the center-of-mass energy $E_{C.M.}$ in MeV.
Solid line for Re $a_{\eta N}$ and dotted line for Im $a_{\eta N}$.}
\label{fig1.}
\end{figure}

\newpage

\begin{figure}[ht]
\includegraphics{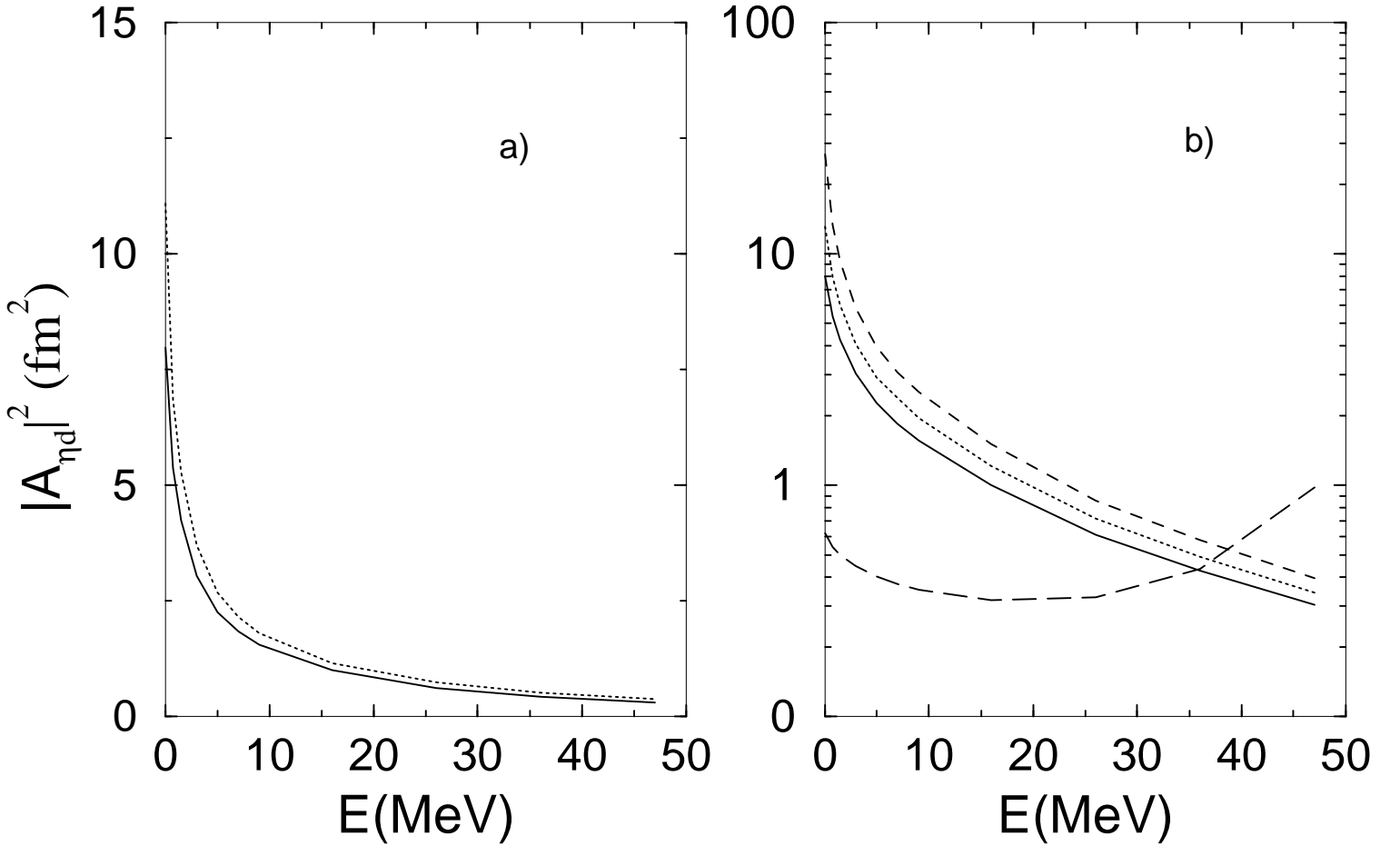}
\caption{ $|A_{\eta d}(E)|^2$ in fm$^2$ as a function of $E$(MeV)  the $\eta-d$
kinetic energy  in the center-of-mass system .
Figure a) shows the results for Re $a_{\eta d}$=0.75 fm but with two
different NN potentials : solid line -- Yamaguchi II and the dashed line
Yamaguchi I.
Figure b) shows the results for Yamaguchi II and Re $a_{\eta N}$ = 0.75 fm
(solid line), 0.87 fm (dotted line), 1.05 fm
(dashed line) and 0.21 fm (long-dashed line) from 
\protect\cite{Kmatrix}\protect\cite{Kmatrix1}. }
\label{fig2ab.}
\end{figure}

\newpage

\begin{figure}[ht]
\includegraphics{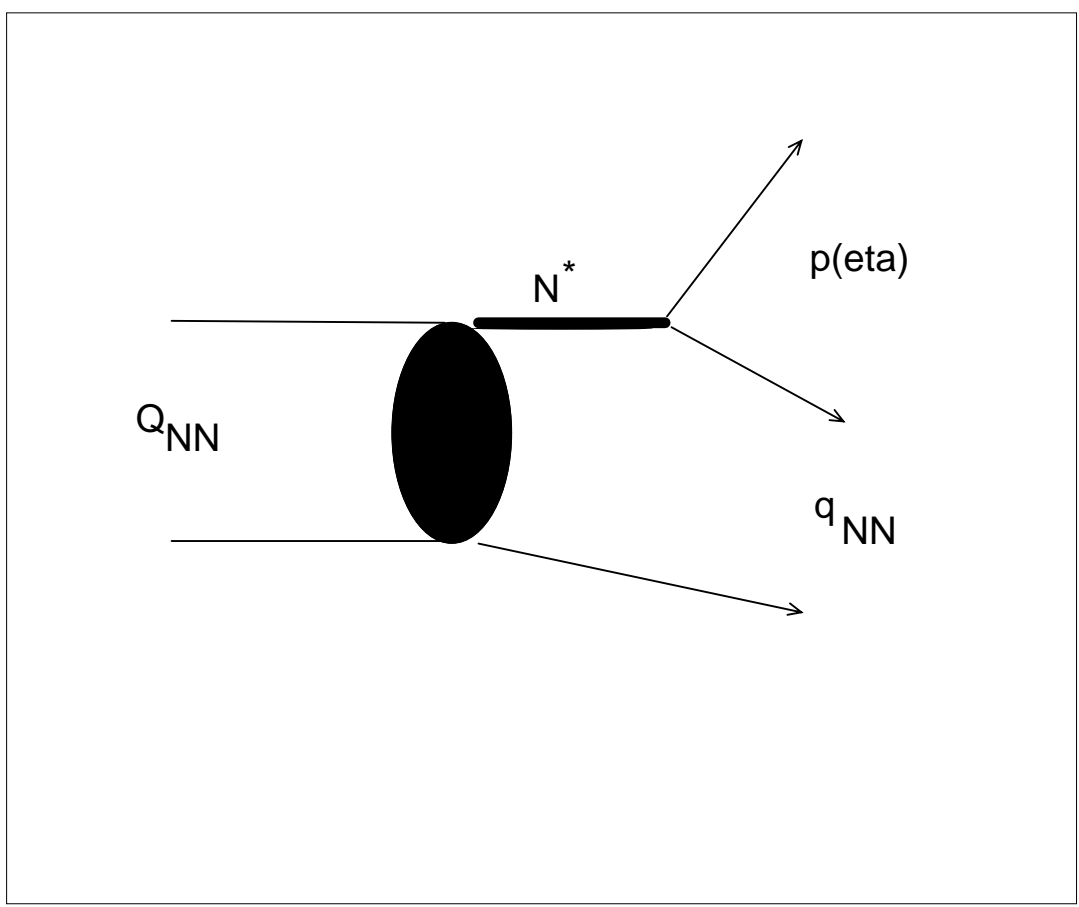}
\caption{ The $\eta$-formation reaction}
\label{fig3}
\end{figure}

\newpage

\begin{figure}[ht]
\includegraphics{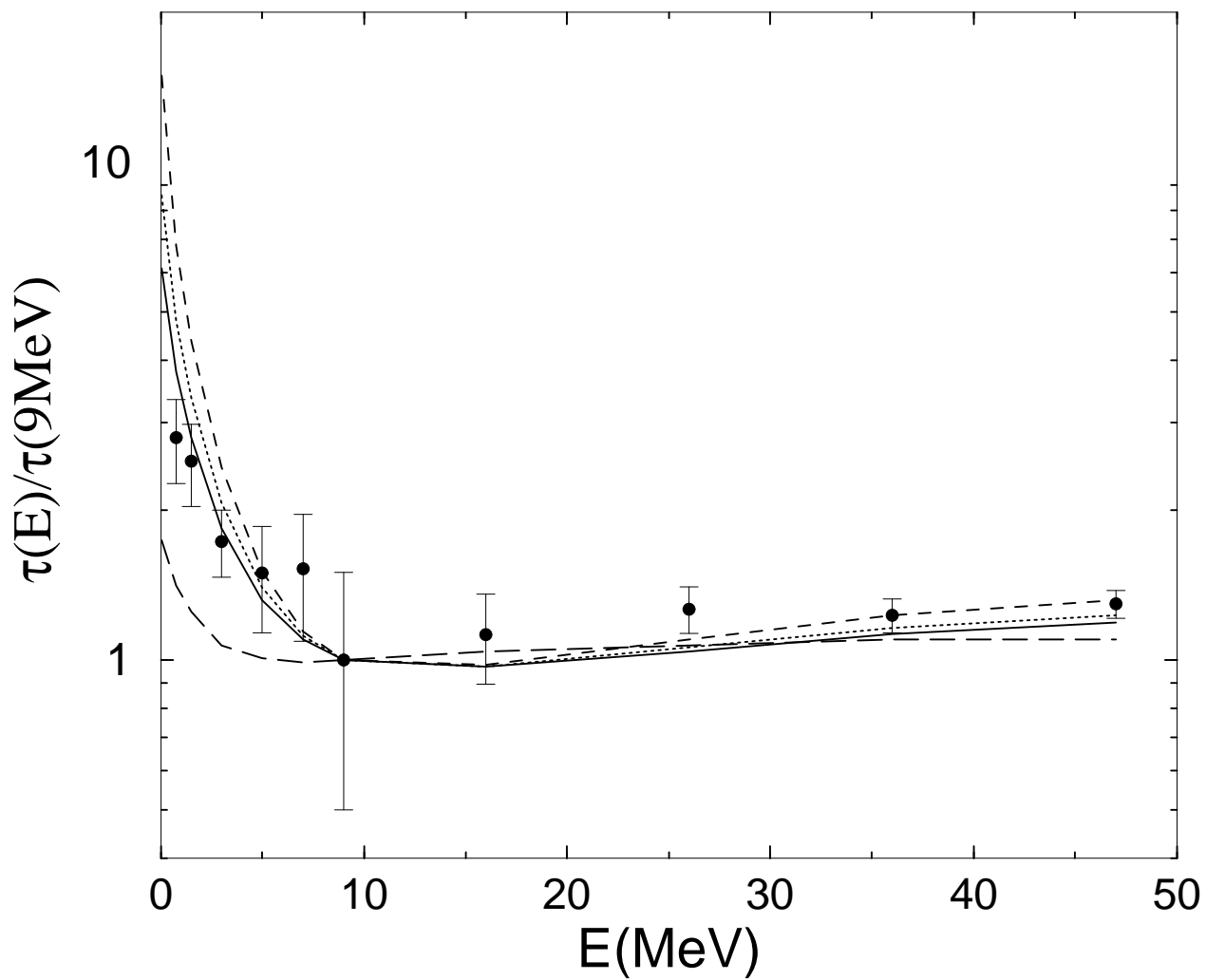}
\caption{ The reduced cross section
$\tau=\sigma(pn\rightarrow \eta d)/p_{\eta}(E)$ from Ref.\protect\cite{pan97}
normalised to unity at 9 MeV. The theoretical equivalent
$|F_{\eta d}(E)|^2/|F_{\eta d}(9 \  {\rm MeV})|^2$ is calculated for
Re $a_{\eta N}$ = 0.75 fm (solid line), 0.87 fm (dotted line), 1.05 fm
(dashed line) and 0.21 fm (long-dashed line) from 
\protect\cite{Kmatrix}\protect\cite{Kmatrix1}.}
\label{fig4}
\end{figure}

\end{document}